\documentclass[letterpaper, 10 pt, conference]{ieeeconf}

\IEEEoverridecommandlockouts

\usepackage{graphics} 
\usepackage{epsfig} 
\usepackage{times, bm} 
\usepackage{amsmath} 
\usepackage{amssymb} 
\usepackage{hyperref}
\usepackage[font=small,labelfont=bf]{caption}
\usepackage{float}
\usepackage{dblfloatfix}
\usepackage{cite}
\usepackage{multirow}

\title{\LARGE \bf
Accelerated MRI with Deep Linear Convolutional Transform Learning
}

\usepackage[]{color-edits}

\author{Hongyi Gu$^{1,2}$,
        Burhaneddin Yaman$^{1,2}$,
        Steen Moeller$^{2}$,
        Il Yong Chun$^{3}$
        and Mehmet Ak\c{c}akaya$^{1,2}$
        \vspace{-0.4cm}
\thanks{$^{1}$Department of Electrical and Computer Engineering, $^{2}$Center for Magnetic Resonance Research, University of Minnesota, Minneapolis, MN, USA, and $^{3}$EEE \& AI, Sungkyunkwan University, Seoul, South Korea.  {\tt\small e-mails: \{gu003818@umn, yaman013@umn, moell018@umn, iychun@skku,
akcakaya@umn\}.edu}}}

\begin{document}
\maketitle
\thispagestyle{empty}
\pagestyle{empty}

\begin{abstract}
Recent studies show that deep learning (DL) based MRI reconstruction outperforms
conventional methods, such as parallel imaging and compressed sensing (CS), in multiple applications. Unlike CS that is typically implemented with pre-determined linear representations for regularization, DL inherently uses a non-linear representation learned from a large database. Another line of work uses transform learning (TL) to bridge the gap between these two approaches by learning linear representations from data.
In this work, we combine ideas from CS, TL and DL reconstructions to learn deep linear convolutional transforms as part of an algorithm unrolling approach. 
Using end-to-end training, our results show that the proposed technique can reconstruct MR images to a level comparable to DL methods, while supporting uniform undersampling patterns unlike conventional CS methods. 
Our proposed method relies on convex sparse image reconstruction with linear representation at inference time, which may be beneficial for characterizing robustness, stability and generalizability.
\newline
\end{abstract}

\section{INTRODUCTION}
\label{sec:introduction}
Conventional accelerated MRI techniques, such as parallel imaging \cite{cgsense} and compressed sensing (CS) \cite{lustig} 
are used extensively in practice, 
but their performance at high acceleration rates is impaired by noise amplification and residual aliasing artifacts. Recently, many studies have demonstrated deep learning (DL) methods show outstanding quality for accelerated MRI
\cite{WangDLMRI,Schlemper,Hammernik,Hemant,Knoll_SPM}. Among DL methods, physics-guided DL (PG-DL) methods that unroll conventional optimization algorithms have been popular \cite{Hammernik, Knoll_SPM, Hemant, Hosseini_JSTSP}. While CS often uses a pre-specified linear representation of images for regularization, PG-DL methods utilize a sophisticated non-linear representation for regularization that is implicitly learned through neural networks. Transfer learning (TL) is another line of work that aims to bridge the gap between these two approaches, where a linear representation is learned from data \cite{TL_Bresler}. Nonetheless, DL methods use large databases, advanced optimization methods, and a large number of parameters, in contrast to the hand-tuning of two or three parameters in CS, or the more traditional optimization strategies employed in TL. On the other hand, sparse processing of linear representations used in CS/TL is more amenable to a theoretical characterization, and may provide a clearer understanding of robustness/stability \cite{JongReviewCS}. Recent literature has aimed to address this gap by incorporating modern data science tools for improving linear representations of MR images. In \cite{Ciuciu}, learning of linear transforms in the context of denoising was explored in a data-driven manner. Another line of work has revisited conventional $\ell_{1}$-wavelet CS for accelerated MRI using state-of-the-art data science tools \cite{Hongyi_ISMRM, Hongyi_EMBC}.

In this work, we go beyond conventional CS by combining ideas from TL \cite{TL_Bresler,Conv_op_learn} and data science tools from PG-DL. Using an $\ell_1$-regularized analysis formulation \cite{Candes_CS_Dictionary}, we unroll an ADMM algorithm and train it end-to-end, using learnable linear convolutional transforms with different receptive fields. The linear representation is designed to be overcomplete, and implemented using cascades of convolution operations for an improved optimization landscape \cite{irani}. The filter coefficients are learned in end-to-end training, along with the $\ell_1$ soft-thresholding parameters.
Results show that the gap in reconstruction performance between the proposed model and more advanced PG-DL methods is minimal; while the proposed method is also able to reliably reconstruct datasets undersampled with uniform undersampling, unlike conventional CS. Thus, the proposed method enables convex sparse image reconstruction using a linear representation at inference time.

\begin{figure*}[t]
 \begin{center}
          \includegraphics[trim={0 0 0 0},clip, width=6 in]{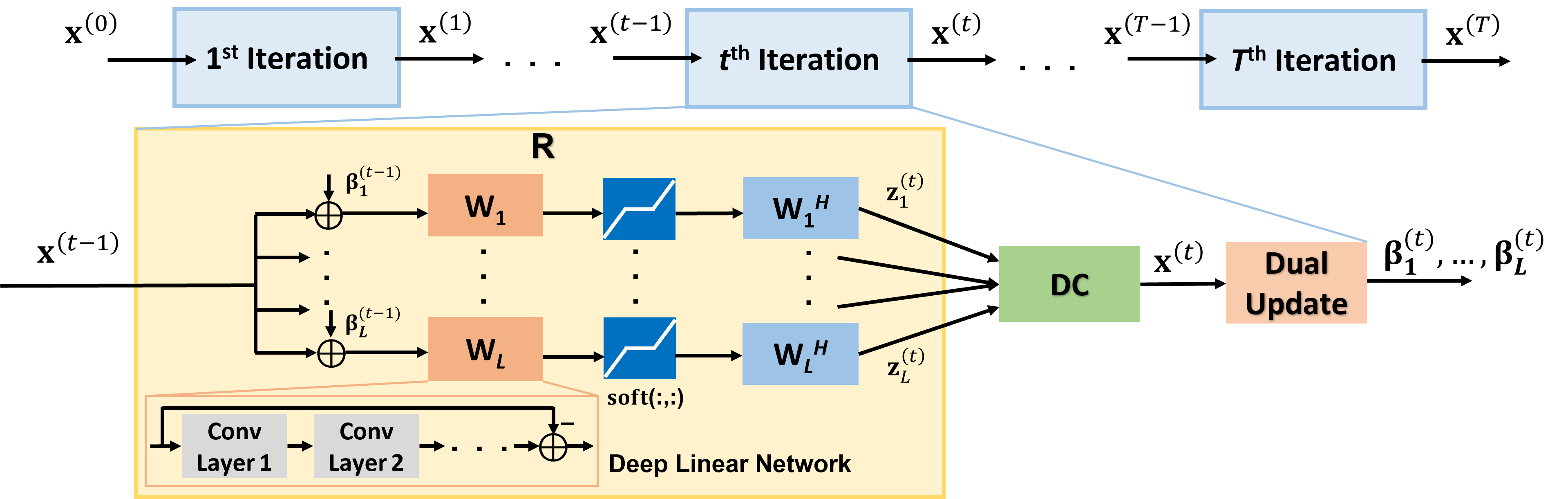}
     \end{center}
      \vspace{-.35cm}
  	\caption{Schematic of the unrolled ADMM for physics-guided reconstruction with deep linear convolutional transform learning (DLC-TL). An unrolled iteration of ADMM consists of a regularizer (R) with linear transforms $\mathbf{W}_{l}$, data consistency (DC) and dual update. All learnable parameters are shared across unrolled iterations. Each ${\bf W}_l$ is encoded as a deep linear network with cascades of convolutional (Conv) layers. Similar to ResNets, a skip connection with inverted sign from the input of each ${\bf W}_l$ is added to its outputs.}
  	\label{fig:ADMM_schematic}
  	\vspace{-.35cm}
\end{figure*}

\section{MATERIALS AND METHODS}

\label{sec:materials and methods}

\subsection{Inverse Problem for Accelerated MRI}
The inverse problem for accelerated MRI is given as 
\begin{equation}
\hat{\mathbf{x}}=\arg \min_{\bf x} \frac{1}{2}\|\mathbf{y}-\mathbf{E} \mathbf{x}\|_{2}^{2}+\cal{R}(\mathbf{x})
\label{eqn.2}
\end{equation}
 where $\mathbf{x}$ is the image of interest, $\mathbf{y}$ is the undersampled multi-coil k-space, $\mathbf{E}$ is the multi-coil encoding operator \cite{cgsense},  $\cal{R}(\mathbf{x})$ is a regularizer. Note the $\|\mathbf{y}-\mathbf{E} \mathbf{x}\|_{2}^{2}$ term enforces data consistency (DC).
The regularizer is chosen as a Tikhonov term in parallel imaging \cite{cgsense} or a sparsity-promoting term in CS \cite{lustig}. In particular, for CS, $\cal{R}(\mathbf{x})$ is usually a weighted $\ell_{1}$-norm of transform coefficients,
i.e. $\mathcal{R}(\mathbf{x})= {\lambda}\left\|\mathbf{W} \mathbf{x}\right\|_{1}$, where $\mathbf{W}$ is a pre-specified linear transform, such as a discrete wavelet transform \cite{lustig}.
The resulting convex objective function is solved via an iterative optimization algorithm \cite{fessler_SPM}, using either hand-tuned or optimized parameters \cite{Hongyi_EMBC}.

\vspace{-.05cm}

\subsection{Transform Learning for MRI Reconstruction}
While conventional CS methods typically adopt pre-specified and commonly available linear sparsifying transforms for $\ell_1$ regularization, TL methods have become popular for their abilities to learn linear representations for reconstruction \cite{TL_Bresler}. Transform learning for MRI solves (\ref{eqn.2}), where $\cal{R}(\mathbf{x})$ is a TL-based regularizer. The regularizer can be either prelearned from the dataset, or learned at the same time with the reconstruction process \cite{TL_Bresler,BCS}. 
While there are a number of variations of the TL methods (please see \cite{TL_Bresler} and the references therein), the general problem can be formulated as
\begin{align}
    \arg \min_{\bf x, W} ||{\bf y - Ex}||_2^2 + \lambda {\cal R}_{s}({\bf Wx}) + \tau {\cal R}_{t}({\bf W}),
\end{align}
where ${\cal R}_{s}$ is a regularizer on the transform-domain signal ${\bf Wx}$, e.g. $\ell_p$ norm with $p \leq 1$; and ${\cal R}_{t}$ is a regularizer on the transform itself to avoid the trivially sparsifying solution ${\bf W = 0}$. Examples of the latter include ${\cal R}_{t}({\bf W}) =\frac12 \|\mathbf{W}\|_{F}^{2}- \log |\operatorname{det} \mathbf{W}|$, which enforces non-degenerate solutions along with scaling constraints, or explicitly enforcing a unitary transform/tight frame condition \cite{TL_Bresler, Conv_op_learn}. Notably, optimization is coupled over ${\bf x}$ and ${\bf W}$, and run until convergence. 

\vspace{-.05cm}

\subsection{PG-DL Reconstruction in MRI}

In PG-DL reconstruction, the inverse problem is usually solved by unrolling an iterative optimization algorithm for a fixed number of iterations \cite{LeCun, Monga}, which alternates between regularizer and DC units. The regularizer in PG-DL is implemented implicitly via neural networks, while the DC unit is solved linearly with methods such as gradient descent or conjugate gradient \cite{Hemant}. The network is trained end-to-end as:
\begin{equation}
    \min_{\bm \theta} \frac1N \sum_{n=1}^{N} \mathcal{L}\Big( {\bf x}_{\textrm{ref}}^n, \: f({\bf y}^n, {\bf E}^n; {\bm \theta})\Big),\label{Eq:2.1_3}
\end{equation}
where ${\bf x}_{\textrm{ref}}^n$ denotes the fully-sampled reference image of the $n^\textrm{th}$ subject, $f({\bf y}^n, {\bf E}^n; {\bm \theta})$ denotes network output of the unrolled network with parameters ${\bm \theta}$ of the $n^\textrm{th}$ subject, $N$ is the number of datasets in the training database, and $\mathcal{L}(\cdot, \cdot)$ is a loss function between the network output and the reference. $\mathcal{L}(\cdot, \cdot)$ is commonly chosen as $\ell_2$ norm, $\ell_1$ norm, mixed norms and perception-based loss \cite{Knoll_SPM,Seitzer}.

\begin{table*}[!b]
\centering
\begin{tabular}{ |p{4.5cm}||p{1.6cm}|p{1.6cm}|p{1.6cm}|p{1.6cm}|p{1.6cm}|p{1.6cm}|  }
 \hline{}Convolutional Transforms
 &\hfil $\mathbf{W}_{1}$ &\hfil $\mathbf{W}_{2}$ &\hfil $\mathbf{W}_{3}$ &\hfil $\mathbf{W}_{4}$ &\hfil $\mathbf{W}_{5}$ &\hfil $\mathbf{W}_{6}$\\
 \hline{} 
 Filter size&
 \multicolumn{3}{|c|}{$3 \times 3$}
 &
 \multicolumn{3}{|c|}{$5 \times 5$}
 \\
 \hline{}
 Number of cascasdes& 
 \hfil2 &
 \multicolumn{2}{|c|}{3}&
 \hfil2&
 \multicolumn{2}{|c|}{3}\\
 \hline
 Dilation rate&
 \multicolumn{2}{|c|}{1}&
 \hfil2&
 \multicolumn{2}{|c|}{1}&
 \hfil2\\
 \hline
 Receptive field size&
 \hfil $5 \times 5$&
 \hfil $7 \times 7$&
 \hfil $11 \times 11$&
 \hfil $9 \times 9$&
 \hfil $13 \times 13$&
 \hfil $21 \times 21$\\
 \hline
 Output size in the cascade&
 \multicolumn{6}{|c|}{28}\\
 \hline
 Parameter number &
 \hfil 7,308&
 \multicolumn{2}{|c|}{14,364}&
 \hfil 20,300&
 \multicolumn{2}{|c|}{39,900}\\
 \hline
\end{tabular}
 \caption{Specifications of each linear convolutional transform.}
 \label{fig:Impl_detail}
\end{table*}

\vspace{-.05cm}

\begin{figure*}[t]
 \renewcommand{\thefigure}{2}
 \begin{center}
          \includegraphics[trim={0 0 0 0},clip, width=6 in]{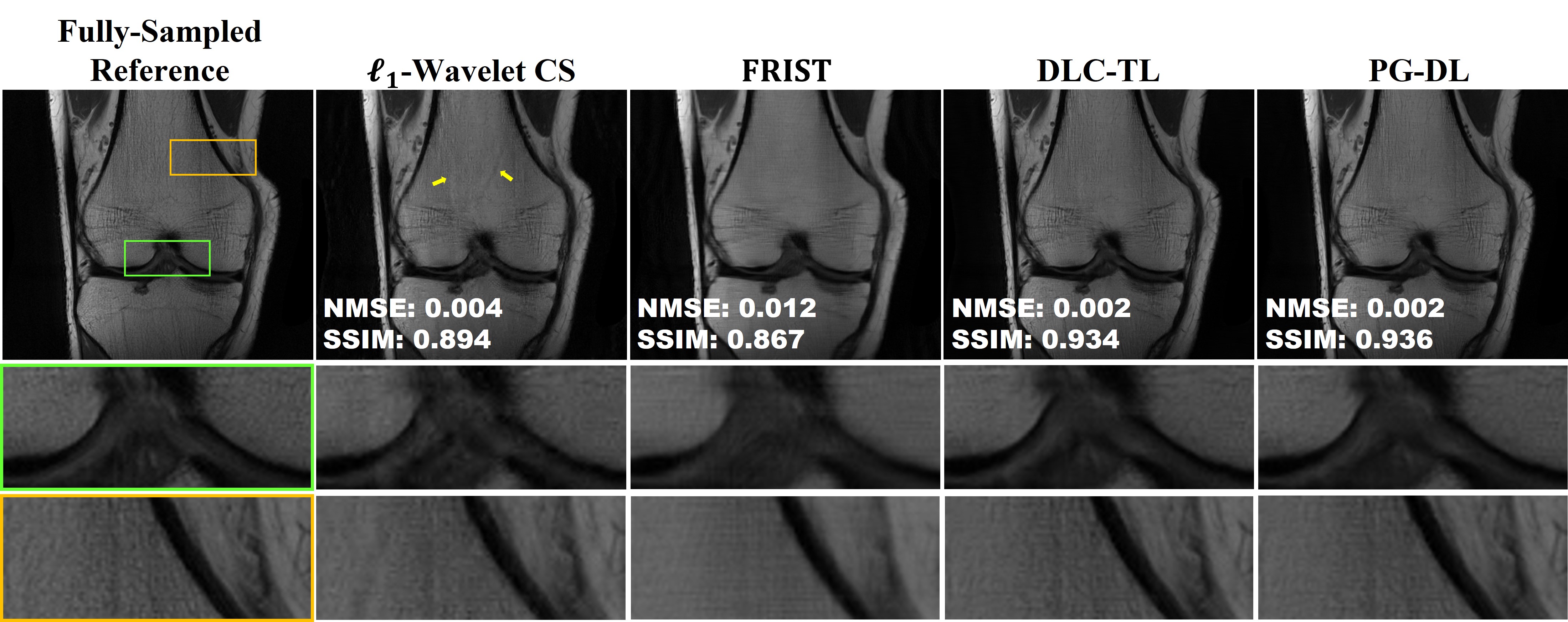}
     \end{center}
      \vspace{-.35cm}

  	\caption{A representative slice from coronal PD knee MRI, reconstructed using $\ell_1$-wavelet CS, FRIST, proposed DLC-TL, and PG-DL. $\ell_1$-wavelet CS exhibits some residual aliasing artifacts. FRIST is able to suppress these, but shows visible blurring. Proposed DLC-TL and PG-DL reduce both aliasing and blurring, with DLC-TL performing closely to PG-DL.}
  	\label{fig:PD_result}
  	\vspace{-.35cm}
\end{figure*}

\subsection{Proposed Learning of Physics-Guided Reconstruction with Deep Linear Convolutional Transforms}

We consider an objective function related to (\ref{eqn.2}):
\begin{equation}
\hat{\mathbf{x}}=\arg \min_{\bf x} \frac{1}{2}\|\mathbf{y}-\mathbf{E} \mathbf{x}\|_{2}^{2}+\sum_{l =1}^L \lambda_l ||{\bf W}_l {\bf x}||_1,
\label{eqn.obj}
\end{equation}
where $\{{\bf W}_l\}$ are linear sparsifying transforms. This corresponds to the analysis formulation of a weighted $\ell_1$-Regularized CS reconstruction problem. Our high-level aim is to learn $\{{\bf W}_l, \lambda_l\}_{l=1}^L$ from a training database, and solve (\ref{eqn.obj}) at inference time using these pre-learned parameters. 
To this end, we first unroll ADMM algorithm to solve (\ref{eqn.obj})
\begin{subequations}
\begin{align}
& \mathbf{x}^{(t+1)}=\bigg(\mathbf{E}^{H} \mathbf{E}+\sum_{l=1}^{L} \rho_{l} \mathbf{I}\bigg)^{-1} \nonumber \\
&\quad \quad \quad \quad \quad \quad \bigg(\mathbf{E}^{H} \mathbf{y}+\sum_{l=1}^{L} \rho_{l} \mathbf{W}_{l}^{H}\left(\mathbf{z}_{l}^{(t)}-\boldsymbol{\beta}_{l}^{(t)}\right)\bigg) \label{ADMM:(a)}
\\
& \mathbf{z}_{l}^{(t+1)}=\textrm{soft}\left(\mathbf{W}_{l} \mathbf{x}^{(t+1)}+\boldsymbol{\beta}_{l}^{(t)} ; \lambda_{l} / \rho_{l}\right) \label{ADMM:(b)}
\\
& \boldsymbol{\beta}_{l}^{(t+1)}=\boldsymbol{\beta}_{l}^{(t)}+\eta_{l}\left(\mathbf{W}_{l} \mathbf{x}^{(t+1)}-\mathbf{z}_{l}^{(t+1)}\right) \label{ADMM:(c)}
\end{align}
\end{subequations}
where $\mathbf{z}_{l}$ are auxiliary variables in the linear transform domain, ${\bm \beta}_{l}$ are dual variables, $\textrm{soft}\left(\cdot; \lambda_{l} / \rho_{l}\right)$ is the $\ell_1$ soft-thresholding operator parameterized by $\lambda_{l} / \rho_{l}$, and $t$ denotes the iteration count. In PG-DL techniques, the algorithm is unrolled and optimized with the $\mathbf{z}_{l}$ updates replaced by CNNs.
In this work, we use linear convolutional operators \cite{Conv_op_learn} for ${\bf W}_l$ with the following loss function:
\begin{align}
    &\min_{{\bm \theta} \triangleq \{{\bf W}_l, \rho_l, \lambda_l, \eta_l \}_{l=1}^L} \frac1N \sum_{n=1}^{N} \Big(\mathcal{L}\Big( {\bf x}_{\textrm{ref}}^n, \: f({\bf y}^n, {\bf E}^n; {\bm \theta})\Big) \nonumber \\
    &\quad \quad \quad \quad \quad+ \epsilon\sum_{l=1}^{L} \|\mathbf{W}_{l}^{H}\mathbf{W}_{l}{\bf x}_{\textrm{ref}}^n- {\bf x}_{\textrm{ref}}^n\|_{2}/ \|{\bf x}_{\textrm{ref}}^n\|_{2}\Big),
\end{align}
where a normalized $\ell_{1}$-$\ell_{2}$ is used for ${\cal L}(\cdot,\cdot)$ \cite{yaman_SSDU_MRM}, $\epsilon$ is a constant weight, and the second term ensures that the learned transforms have properties similar to tight frames over the training set, though an explicit tight frame condition \cite{Conv_op_learn} has not yet been investigated. 

Our final contribution is to encode each ${\bf W}_l$ as a deep linear network \cite{saxe2014exact}, featuring cascades of convolutional layers, in order to leverage the large-scale optimization algorithms used in DL applications, and the associated optimization landscape that tends to have multiple good-performing local minima \cite{loss_surfaces}. While this does not change the expressiveness of the linear convolutional operator, the optimization has different properties, notably the presence of infinitely many valid solutions for the training objective and faster convergence \cite{irani}. Thus, each $\mathbf{W}_{l}$ is implemented as multiple cascades of convolutional layers $\{\mathbf{W}_{l,q}\}_q$. Each $\mathbf{W}_{l}$ is designed to have distinct receptive field sizes, with the aim of mapping the input image to distinct feature spaces for better regularization. We refer to this overall approach as deep linear convolutional transform learning (DLC-TL). Finally, we incorporate the idea of residual learning \cite{He_2016_CVPR}, and add a skip connection path from the input of each ${\bf W}_l$ to its outputs. Specifically, the input scaled by $1/C$, where $C$ is the output size in the cascade (as in Table \ref{fig:Impl_detail}), is subtracted from the output of ${\bf W}_l$. A schematic of the overall strategy is depicted in Figure \ref{fig:ADMM_schematic}.

\subsection{Imaging Data}
Fully-sampled coronal proton density (PD) and PD with fat-suppression (PD-FS) knee data obtained from the NYU-fastMRI database \cite{fastmri} were used throughout the experiments. Relevant imaging parameters were: matrix size = $320 \times 368$, in-plane resolution = $0.49 \times 0.44 \textrm{ mm}^2$, slice thickness = $3$ mm. The datasets were retrospectively under-sampled with a uniform mask ($R = 4$ with 24 ACS lines). Training was performed on 300 slices from 10 different subjects. Testing was performed on all slices from 10 different subjects. Coil sensitivity maps were generated using ESPIRiT.

\subsection{Implementation Details}
For our proposed DLC-TL, $L = 6$ linear transforms were used. $\mathbf{x}$ is transformed by each $\mathbf{W}_{l}$ with real and imaginary components processed separately with the same networks. $\mathbf{W}_{l}$ either uses two cascades, i.e. $\mathbf{W}_{l} = \mathbf{W}_{l,1}\mathbf{W}_{l,2}$, or three cascades, i.e. $\mathbf{W}_{l} = \mathbf{W}_{l,1}\mathbf{W}_{l,2}\mathbf{W}_{l,3}$. Table \ref{fig:Impl_detail} summarizes the network structures being implemented for $\{\mathbf{W}_{l}\}_{l=1}^L$. Note that for each cascade, different input and output pairs requires learning a distinct convolutional filter. Including learnable $\{\rho_{l}, \lambda_l, \eta_l \}_{l=1}^L$, the total parameter number for the proposed model is 136,154.

ADMM algorithm was unrolled for $T=10$. DC subproblem was solved using conjugate gradient \cite{Hemant} with 5 iterations and warm-start. All parameters were randomly initialized. Adam optimizer with learning rate $5\cdot 10^{-4}$ was used. Supervised training was performed on TensorFlow in Python, over 100 epochs, with a batch size of 1.

A PG-DL approach implemented using the same ADMM unrolling except for using a ResNet-based regularizer unit was used for comparison, which is similar to ADMM-CS Net \cite{ADMM-CS}, but modified for multi-coil implementation, and which has been used in multiple recent MRI studies successfully \cite{yaman_SSDU_MRM,Hosseini_JSTSP}. The PG-DL approach has a total of 592,130 learnable parameters. Note this constitutes a head-to-head comparison, with the only difference being in the $\mathcal{R}(\mathbf{x})$ term, where our proposed approach employs learnable deep convolutional operations for solving a convex problem, while PG-DL uses a CNN for implicit regularization. An $\ell_1$-wavelet CS approach, and a state-of-the-art TL approach, Flipping and Rotation Invariant Sparsifying Transform (FRIST) \cite{wen2017frist}, were also used for comparison. All results were quantitatively compared using NMSE and SSIM.

\begin{figure*}[t]
 \renewcommand{\thefigure}{3}
 \begin{center}
          \includegraphics[trim={0 0 0 0},clip, width=6 in]{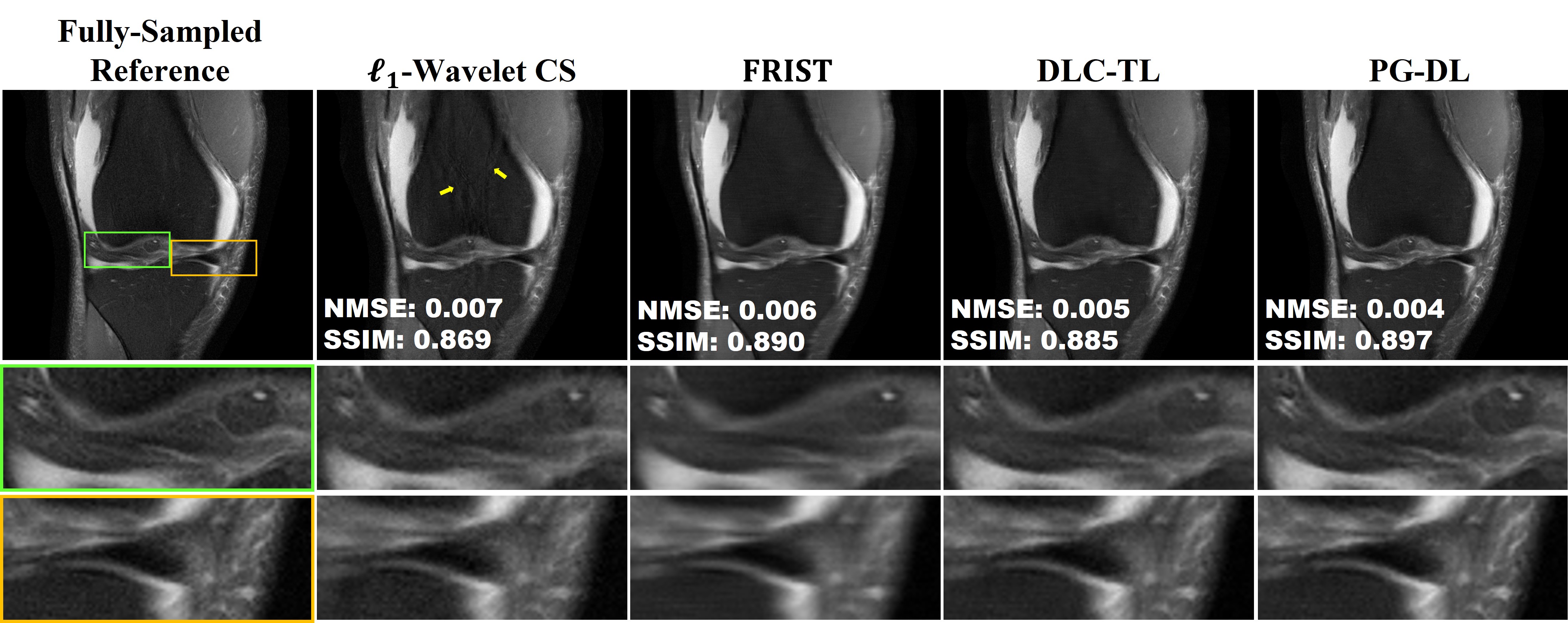}
     \end{center}
      \vspace{-.35cm}

  	\caption{A representative coronal PD-FS knee slice, reconstructed using $\ell_1$-wavelet CS, FRIST, DLC-TL, and PG-DL. Residual aliasing remains in $\ell_1$-wavelet CS, while FRIST is adversely affected by visible blurring. DLC-TL and PG-DL have significantly fewer artifacts, with both methods leading to visibly similar reconstructions.}
  	\label{fig:PDFS_result}
  	\vspace{-.35cm}
\end{figure*}

\begin{figure*}[b]
 \renewcommand{\thefigure}{4}
 \begin{center}
          \includegraphics[trim={0 0 0 0},clip, width=6 in]{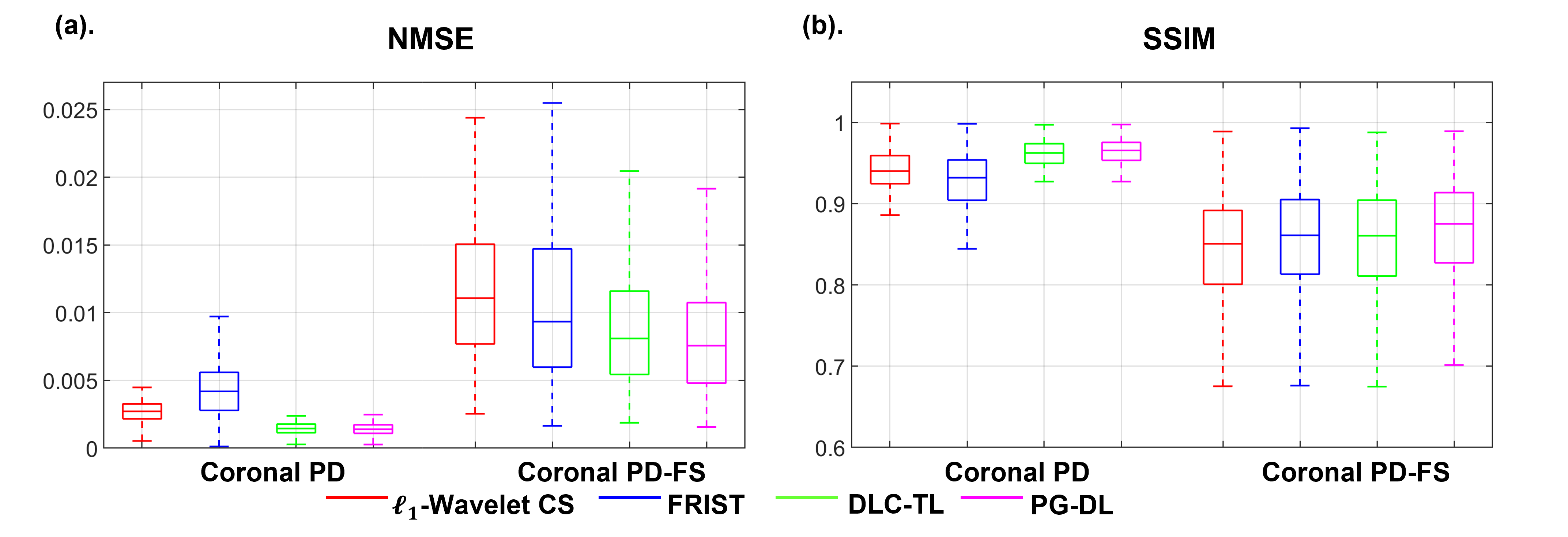}
     \end{center}
      \vspace{-.35cm}

  	\caption{Boxplots showing the median and interquartile range (25th-75th percentile) of (a) NMSE and (b) SSIM metrics on test slices from 10 subjects for coronal PD and PD-FS datasets. Both DLC-TL and PG-DL outperform $\ell_{1}$-wavelet CS and FRIST, while having comparable quantitative metrics.}
  	\label{fig:Metrics}
  	\vspace{-.35cm}
\end{figure*}

\section{RESULTS}
\label{sec:results}

Figure \ref{fig:PD_result} and \ref{fig:PDFS_result} show representative reconstruction from coronal PD and PD-FS knee MRI, respectively. The proposed DLC-TL and PG-DL remove aliasing artifacts that are present in $\ell_1$-wavelet CS. Note the CS reconstruction typically works with random undersampling patterns, thus its performance is degraded in the uniform undersampling pattern case considered in this work. FRIST is able to reduce residual aliasing, but suffers from visible blurring. Sharpness is maintained by DLC-TL and PG-DL, compared to blurring that is more apparent with $\ell_1$-wavelet CS and FRIST, especially for coronal PD-FS data.

Figure \ref{fig:Metrics} depicts the quantitative metrics over all test datasets, showing  the median and interquartile ranges (25th-75th percentile) of NMSE and SSIM metrics. Both DLC-TL and PG-DL outperform $\ell_{1}$-wavelet CS and FRIST, while having comparable quantitative metrics.

\section{DISCUSSION AND CONCLUSION}
\label{sec:discussion and conclusion}
In this study, we proposed a combination of ideas from TL, CS and PG-DL literatures to learn deep linear convolutional transform for MRI reconstruction. Both the state-of-the-art PG-DL and the proposed DLC-TL outperform $\ell_1$-wavelet CS, which suffers from aliasing artifacts for uniform undersampling case as expected, and a state-of-the-art TL method, FRIST, which exhibits visible blurring. While PG-DL quantitatively outperforms our approach, the performance gap is small, with $<0.0006$ difference in NMSE and $<0.015$ difference in SSIM between the two methods. Furthermore, our proposed method leads to a linear representation, is more interpretable, and enables convex optimization for the CS-based inverse problem at inference time.

Further work on enforcing an explicit tight frame condition \cite{Conv_op_learn} for DLC-TL is warranted for further performance gains, along with a principled extensive optimization of the deep linear networks used for $\{{\bf W}_l\}_{l=1}^L$.

\section{ACKNOWLEDGEMENTS}
This work was partially supported by NIH R01HL153146, NIH P41EB027061, NIH U01EB025144; NSF CAREER CCF-1651825.
\label{sec:acknowledgements}

\bibliographystyle{IEEEbib}
\bibliography{reference}
\end{document}